\documentclass[11pt,a4paper]{article}

\usepackage{bm,amsmath,amssymb}
\usepackage{graphicx}
\usepackage{lineno,hyperref}
\usepackage{fancyhdr}
\usepackage[]{algorithm2e}

\newcommand{\kv}{$k$-vector}
\newcommand{\B}[1]{{\bm #1}}
\newcommand{\ds}{\displaystyle}

\fancypagestyle{firststyle}
{
	\fancyhf{}
	\fancyhead[RE,LO]{Applied Mathematics and Computation Vol. 320, pp. 754-768, 2018. \\
		doi: 10.1016/j.amc.2017.10.009}
	\fancyfoot[RE,LO]{10.1016/j.amc.2017.10.009}
	\fancyfoot[LE,RO]{Page \thepage}
}

\begin{document}

\title{Nonlinear Function Inversion using \kv}

\author{David Arnas\thanks{Centro Universitario de la Defensa, Zaragoza, Spain}, Daniele Mortari\thanks{Aerospace Engineering, Texas A\&M University, College Station TX, USA}}

\date{}	

\maketitle{} 	

\thispagestyle{firststyle}

\begin{abstract}
    This work introduces a general numerical technique to invert one dimensional analytic or tabulated nonlinear functions in assigned ranges of interest. The proposed approach is based on an ``optimal'' version of the \kv\ range searching, an ad-hoc modification devised for function inversion. The optimality consists of retrieving always the same number of data ($1,2,\dots$) for a specified searching range to initiate the root solver. This provides flexibility to adapt the technique to a variety of root solvers (e.g., bisection, Newton, etc.), using a specified number of starting points. The proposed method allows to build an inverse function toolbox for a set of specified nonlinear functions. In particular, the method is suitable when intensive inversions of the same function are required. The inversion is extremely fast (almost instantaneous), but it requires a one-time preprocessing effort.
\end{abstract}

\section{Introduction}\label{sec:intro}

Many problems in science involve computing the inverse of nonlinear functions with unknown analytical expressions~\cite{NIST}. Examples of these are: the Airy function~\cite{Mainardi,Jonsson} (in optics, fluid mechanics, elasticity, or quantum physics), the Bessel's integrals~\cite{Jackson, Bailey} (in many scientific problems, as in quantum field, and condense matter theory problems), the Fresnel's integrals~\cite{Feynman, Boersma} (in quantum mechanics and optics), the Dawson's integrals~\cite{Niu, Hu} (in relative hydrodynamics and wave problems), and the error function~\cite{Strecok, Winitzki}. A great number of numerical algorithms have been developed over the years to calculate these functions, while their inverses are computed by classic root finders and proper initial guesses. The key point here is: \emph{how to obtain a very accurate initial guess as fast as possible}? The purpose of this paper is, indeed, the development of an algorithm providing (almost instantaneously) initial guesses with accuracy increasing with the memory available.

The most classic and common root finders are Newton-Raphson, regula-falsi, bisection, secant, and fixed-point. Recently, other general root finders were developed to increase performance, based on Adomian decomposition method~\cite{Abbasbandy1, Abbaoui} and modifications of the Newton-Raphson algorithm~\cite{Neta0, Abbasbandy2}. However, these methodologies only allow to find one root at a time in nonlinear equations, with increasing complexity when multiple roots are to be computed.

In this work, we introduce a general algorithm to invert one-dimensional functions, which is based on generating a first approximation to the root using the \kv\ range searching methodology. Then, the final root is obtained by the application of another root finder such as Newton-Raphson or regula-falsi. The advantages of this methodology over the existing methods is that \emph{it allows to find all the roots of the function at the same time} and that \emph{it can be adapted to each problem in study}. On the other hand, its disadvantage is that it requires to perform an initial preprocessing for each function, making the method suitable when intensive inversions of the same function are required and/or when a toolbox of prescribed inverse functions is needed.

The \kv\ technique has its origin in the spacecraft attitude estimation problem using star trackers, where the identification of the observed stars (Star-ID) is required in order to estimate the spacecraft orientation. This process must be done as fast as possible (to improve the performance of the control system) by an onboard computer with limited capabilities. To solve this problem, a fast range searching technique, called \kv, was developed~\cite{Original, SLA}. Currently, the \kv\ is at the heart of \emph{Pyramid}~\cite{pyramid}, the state-of-the-art of Star-ID algorithms, currently operating in several satellites

The most important property of the \kv\ is that, once the preprocessing effort is done for a given database, \emph{the searching process becomes independent from the database size}, which makes the method specially suitable for large databases. Compared with the most common searching algorithm, the binary-search technique, the \kv\ has a complexity of $\mathcal{O}(3)$, whilst the binary-search technique has a complexity of $\mathcal{O}(2\log_2 n)$, where $n$ is the database size.

The \kv\ is a general range searching technique, particularly suitable for retrieving data from static databases (such as star catalogs)~\cite{Neta}. However, this technique has recently been applied to solve a variety of different problems, such as sampling~\cite{random}, interpolation, and estimation~\cite{Rogers}.

This work focuses on the application of the \kv\ technique to fast function inversion, introducing a general methodology to invert any one-dimensional nonlinear function in specific ranges of interest. In particular, the technique allows to speed the convergence of other root finders, providing an initial root approximation, or for bracketing all the roots of a given function. This study shows how to solve particular problems the technique may deal with, such as singularities of the function, piecewise function definition, and functions with regions with very different derivative values. Moreover, and in order to improve the performance of the function inversion, an \emph{optimal} \kv\ technique is here introduced, which can be applied in combination with any root solver to obtain machine error precision.

The optimal \kv\ is an ad-hoc improvement of the \kv\ methodology, specifically developed for function inversion. It is based on the idea of finding the distribution of points, for a specific function, in order \emph{to always retrieve the same number of elements} to initiate the root finder. This way, the searching algorithm becomes adaptive, by retrieving the initial guess(es) as the selected root solver asks for. One example of this is to get two points bracketing the roots, a property that is required for bisection and regula-falsi algorithms. In case a Newton root solver is selected, the \kv\ approach is able to provide just one point -- the closest to the root -- for the iteration initialization.

Furthermore, when the function to invert is computationally intensive, the inverse process can be performed \emph{with no function evaluations}. The process requires a more intensive preprocessing, producing in general a larger database that will be used to get fast approximations of the roots with no function evaluations. This will reduce the root accuracy but it increases the inverse process speed considerably.

Simple and detailed examples are presented in order to show the procedure and potential of the methodology in a clear manner. These examples include the Gamma function, the Airy function of the first kind, the Bessel integral, Kepler's equation and the Gaussian integral.

\section{Background on the \kv}\label{sec:background}

The range searching problem consists on retrieving, in a database of size $n$, all the elements that are contained in a given interval, $[y_a, \: y_b]$, where $y_a$ and $y_b$ are the lower and upper bounds of the interval, respectively. The most common searching algorithms are based on the \emph{binary search} technique, that has complexity $\mathcal{O}(2\log_2 n)$, and the \emph{search by hashing}, which is fast on the average case while it can be linear in the worst case due to the collision problem\footnote{The collision issue is due to the use of modular arithmetic to assign bins to elements. As a consequence of that, elements with values far from each other may be assigned to the same bin.} affecting the method. The \kv\ range searching technique is as fast as a hashing method (with best case complexity $\mathcal{O}(3)$), which has no collision problem but that may include some of the \emph{closest elements} to the query range\footnote{The expected number of these extraneous elements depends on the nonlinearity of the database.}. However, it requires a very fast preprocessing effort (provided in the appendix), as fast as just reading the sorted database. These properties make the \kv\ particularly suitable for searching in large databases, where the worst case complexity goes asymptotically to $\mathcal{O}(3)$ as more memory is available to store the \kv\ elements~\cite{MAKV}.

\begin{figure}[!h]
	\centering
	{\includegraphics[width=0.47\textwidth]{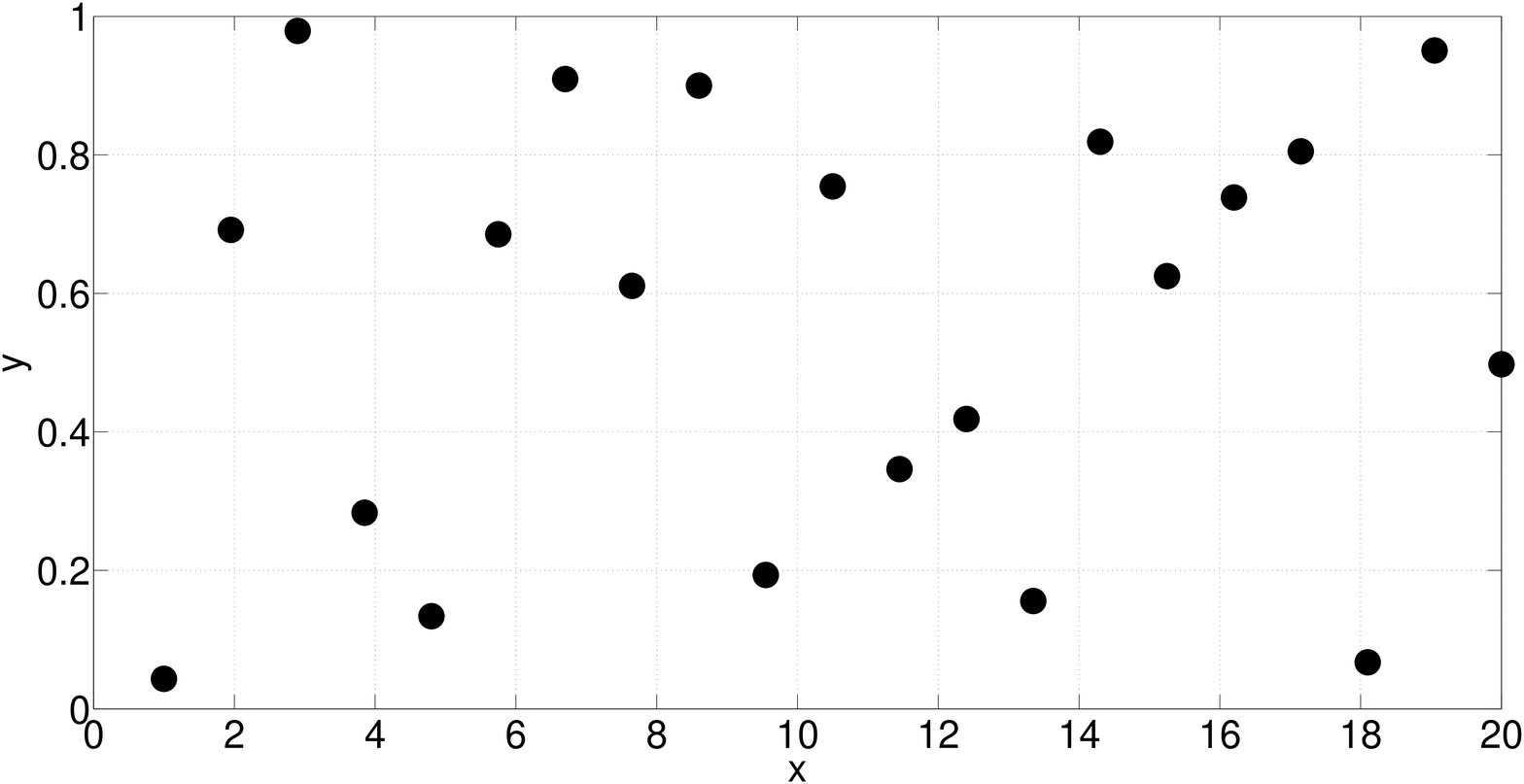}} \hspace{0.5cm}
	\includegraphics[width=0.47\textwidth]{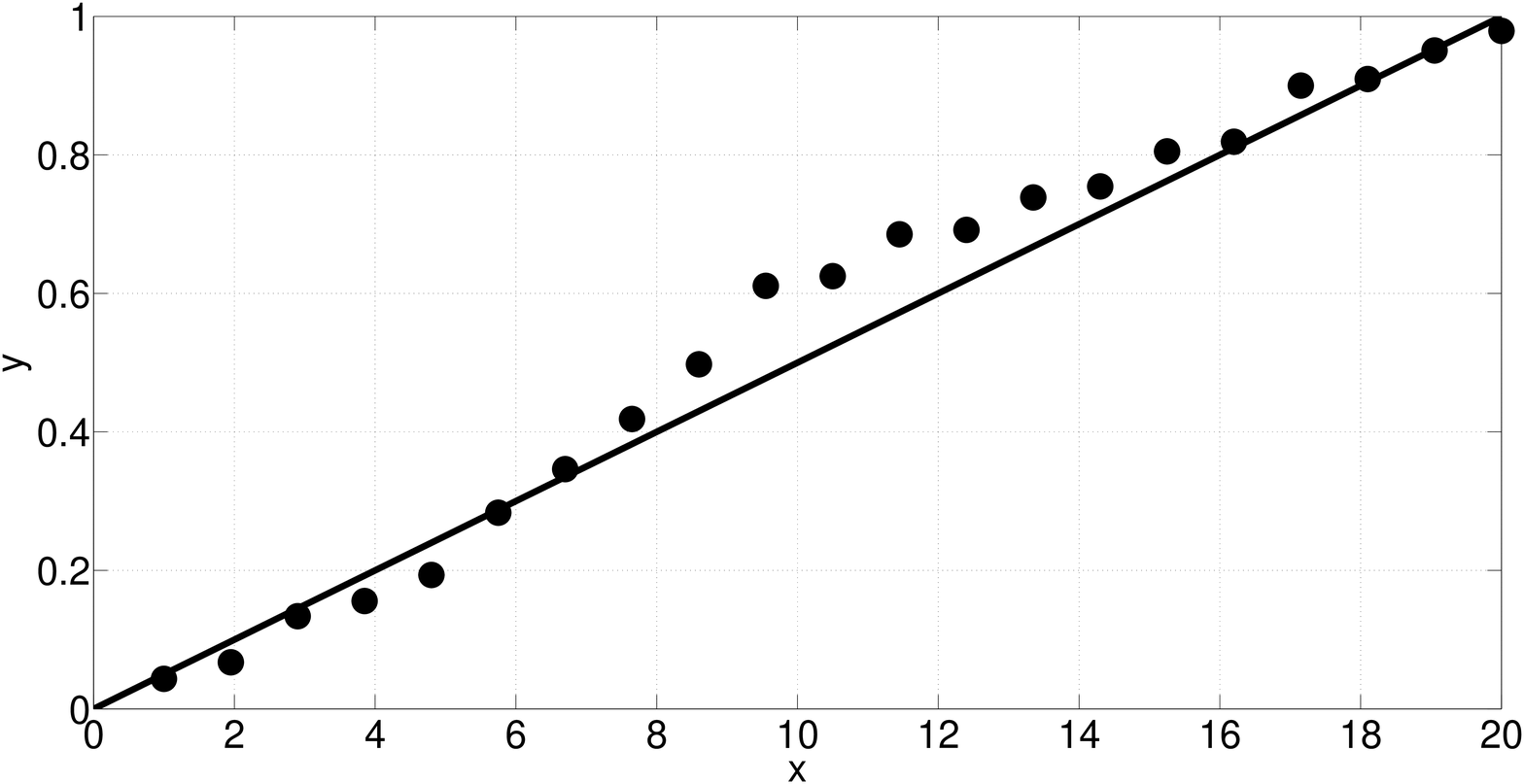}
	\caption{Unsorted random database (left) and \kv\ line and sorted database (right).}
	\label{fig:basics}
\end{figure}

The \kv\ is built using a sorted database ordered in the ascending mode and then generating a vector of indexes containing the information of the number of elements below a mapping function for specific values of the function. In this document, a straight line is assumed to be used as mapping function, since it provides the best speed performance for the methodology. Figure~\ref{fig:basics} shows a random database (left), the sorted database (right), and the \kv\ mapping function (line).

Let $n$ be the number of elements of the database, being $\B{y}(i)$ the $i$-th element, and $\B{s}$ the vector containing the sorted database $\B{y}$, that is:
\begin{equation}
	\B{s}(i) \leq \B{s}(i+1), \quad \forall i\in[1,n-1].
\end{equation}
Therefore, the minimum and maximum values of the database are $y_{\min} = \B{s} (1)$ and $y_{\max} = \B{s} (n)$. Let $\B{I}$ be the sorting indexes vector relating $\B{y}$ and $\B{s}$, that is:
\begin{equation}
	\B{s}(i) = \B{y}(\B{I}(i)) \quad \text{where} \quad i = \{1, 2, \cdots, n\}.
\end{equation}

The \kv\ is a vector of indexes containing information about the nonlinearity of the sorted database, that is, the variation with respect to the \kv\ mapping function (the \kv\ line). In order to include all the elements (given the rounding and machine errors), the \kv\ line must connect the points $[1, y_{\min} - \delta\varepsilon]$, and $[n, y_{\max} + \delta\varepsilon]$, where
\begin{equation}
	\delta\varepsilon = (n - 1) \varepsilon,
\end{equation}
and $\varepsilon$ is the relative machine precision ($2.22\times 10^{-16}$ for double precision arithmetic). Therefore, the \kv\ line equation is,
\begin{equation}\label{kline}
	\B{y}_l(i) = m \, (i - 1) + q = \dfrac{y_{\max} - y_{\min} + 2\delta\varepsilon}{n - 1} (i - 1) + y_{\min} - \delta\varepsilon,
\end{equation}
with
\begin{equation}
	m = \dfrac{y_{\max} - y_{\min} + 2\delta\varepsilon}{n - 1}, \qquad \text{and} \qquad q = y_{\min} - \delta\varepsilon,
\end{equation}
and where $i = \{1, 2, \cdots, n\}$. The line defined in Eq. (\ref{kline}) has the purpose to define a series of reference levels, $\B{y}_l(i)$, that are used to generate the \kv.

The \kv\ ($\B{k}_v$) stores the nonlinearity of the sorted database by counting the number of elements that are below a given level defined by Eq. (\ref{kline}). This is equivalent to $\B{k}_v (i) = j$, where $j$ is the greatest index that fulfills $\B{s} (j) \leq \B{y}_l (i)$, that is:
\begin{equation}
	\B{k}_v (i) = \max\left(\{j \mid \B{s} (j) \leq \B{y}_l (i)\}\right).
\end{equation}
This also implies that $\B{k}_v (1) = 0$ and $\B{k}_v (n) = n$, no matter the database studied, since there cannot be elements below $\B{y}_l (1)$ nor above $\B{y}_l (n)$. Appendix A provides a pseudo-code to generate the \kv.

Once the \kv\ is built, retrieving the elements that are inside the range $[y_a, \, y_b]$ becomes an easy task. Let $k_a$ and $k_b$ be the two indexes that correspond to the range searching boundary $[y_a, \, y_b]$. They are computed using Eq. (\ref{kline}),
\begin{equation}
	k_a = \left\lfloor\dfrac{y_a - q}{m}\right\rfloor + 1, \qquad \text{and} \qquad k_b = \left\lceil\dfrac{y_b - q}{m}\right\rceil,
\end{equation}
where $\lfloor x \rfloor$ is the greatest integer lower than $x$, and $\lceil x \rceil$ is the lowest integer greater than $x$. Using the $k_a$ and $k_b$ indexes is then possible to find the range of elements in $\B{y}$ that are in the interval $[y_a, \: y_b]$. Let $\{k_a:k_b\}$ be the set of integer indexes from $k_a$ to $k_b$, that is, $\{k_a:k_b\} = \{k_a , k_a + 1, \cdots, k_b - 1, k_b\}$. Then, the searched elements are:
\begin{equation}
	\{\B{y}(i) \in \B{y}\mid \B{y}(i) \in [y_a, \: y_b]\} = \B{y}(\B{I}(k_a:k_b)).
\end{equation}
Appendix B provides the pseudo-code to use the \kv and retrieve the elements sought.

Using this procedure it may be possible that some extraneous elements -- \emph{the closest to the searching range} -- will be included in the data retrieved. In particular, since the number of $\B{y}_l(i)$ bins/steps are $(n - 1)$ and the number of elements are $n$, there will be an average of $E_0 = n/(n - 1)$ elements in each $[\B{y}_l (i), \B{y}_l (i+1)]$ bin. This means that the expectation of the number of these extraneous element is $n/(n - 1)$, a value that is close to one for large databases. This happens due to the $50\%$ probability for each of the two external bins to get elements lower than $y_a$ or higher than $y_b$. In the case that these elements (the closest to the $[y_a, y_b]$ range) cannot be tolerated, they may be removed from the retrieved data by two local searches,
\begin{equation}\label{out}
	\B{y}(\B{I}(k_a \rightarrow)) < y_a \qquad \text{and} \qquad \B{y}(\B{I}(\leftarrow k_b)) > y_b,
\end{equation}
by increasing the indexes from $k_a$ and decreasing the indexes from $k_b$ as long as the inequalities in Eq. (\ref{out}) are satisfied. As it can be seen, other than removing an average of $n/(n - 1)$ elements, this technique does not require to perform any kind of search. This fact highlights the most important property of the \kv: \emph{the algorithm complexity is not a function of the database size}.

\section{Finding the roots of a function}\label{sec:roots}

This section shows the procedure to obtain the roots of an assigned nonlinear function, $y = f(x)$, within the ranges $[x_{\min}, \, x_{\max}]$ and $[y_{\min}, \, y_{\max}]$ using the \kv\ technique. The methodology requires a preprocessing effort where the function is discretized, generating a database that is first sorted and then accessed using the \kv. This preprocessing effort must be performed \emph{just once} for each nonlinear function considered. For this reason, the proposed method is suitable to build a toolbox of a set of inverse functions.

First, the process requires to perform a function discretization in $x$ (e.g., with constant step) within the domain of interest. Let $n$ be the number of elements of the database and $\B{x}$ and $\B{y}$ be the vectors containing the database elements, where:
\begin{equation}
	\B{y}(i) = f(\B{x}(i)), \quad \text{with} \quad i = \{1,2,\cdots,n\}.
\end{equation}
Second, the maximum absolute difference, $\delta$, between two consecutive values of $\B{y}$ is computed as:
\begin{equation}
	\delta = \max\left(\left|\B{y}(i + 1) - \B{y}(i)\right|\right) + 4 \varepsilon \quad \text{with} \quad i \in [1,n-1].
\end{equation}
This parameter is important as it allows to define a searching range for the \kv\ methodology in such a way that at least one discrete point is found close to each root. In particular, let $y_r$ be the value of the function to be inverted. Then, the function whose roots are required to be computed is:
\begin{equation}
	f(\{x_r\}) - y_r=0,
\end{equation}
where the set of roots for the value $y_r$ is denoted by $\{x_r\}$. In order to retrieve always at least one point near each root, the minimum searching range must be:
\begin{equation}
	[y_a, \; y_b] = \left[y_r - \dfrac{\delta}{2}, \; y_r + \dfrac{\delta}{2}\right].
\end{equation}
Third, the $\B{y}$ table is sorted in ascending mode and the \kv\ is built. Using the searching range, $[y_a, \; y_b]$, the \kv\ retrieves the two indexes $k_a$ and $k_b$, that are used to generate the vector $\B{k} = \{k_a:k_b\}$, which contains all the indexes of database points near the roots. All points close to the roots are $\{\B{x} (\B{I} (\B{k}))\}$ and $\{\B{y} (\B{I} (\B{k}))\}$. However, these points \emph{are not in order}. To find the number of roots and their locations the set $\{\B{x} (\B{I} (\B{k}))\}$ -- generally small -- needs to be sorted,
\begin{equation}
	\B{x}_s = \{\B{x} (\B{I}_x (\B{I} (\B{k})))\},
\end{equation}
where $\B{I}_x$ is the sorting index vector. The values of the function in these sorted points are,
\begin{equation}
	\B{y}_s = \{\B{y}(\B{I}_x (\B{I}(\B{k})))\}.
\end{equation}
Since the discretization is uniform, the $x$-distance between two consecutive points is:
\begin{equation}
	\delta_x = \dfrac{x_{\max} - x_{\min}}{n - 1}.
\end{equation}
Therefore, a sequential search is performed in $\B{x}_s$ to count the number of different roots obtained. If
\begin{equation}\label{nroot1}
	\B{x}_s (i+1) - \B{x}_s (i) < 1.5 \, \delta_x,
\end{equation}
then the two points with indices $i$ and $i+1$ belong to the same root, otherwise, if
\begin{equation}\label{nroot2}
	\B{x}_s (i+1) - \B{x}_s (i) > 1.5 \, \delta_x,
\end{equation}
they belong to two consecutive roots. This allows to discriminate the roots (by grouping the nearest points to each root) and to count them.

Let $\{x_{rr}\}$ be the set of consecutive points near to a particular root. Then, the closest element to the root, $x_{rr0} \in \{x_{rr}\}$, is
\begin{equation}
	x_{rr0} = \{x_{rri}\in x_{rr} :  |f(x_{rri}) - y_r| = \min_{x_{rrj}\in\{x_{rr}\}} |f(x_{rrj}) - y_r|\}.
\end{equation}
The value of $x_{rr0}$ can be then used as initial guess to find the root by Newton-Raphson iterations
\begin{equation}
	x_{(rr,k+1)} = x_{(rr,k)} - \dfrac{f (x_{(rr,k)})}{f' (x_{(rr,k)})}, \quad \text{with} \quad x_{(rr,0)} = x_{rr0}.
\end{equation}
Since $x_{rr0}$ is really close to the root, the Newton-Raphson iteration process converges with machine error accuracy in one or two iterations, only. However, in order to avoid special situations when the convergence may not be obtained (or even experience divergence), two sanity checks are considered during the iteration process. The first consists of bounding the maximum number of iterations, while the second checks if the updated value remains within a small range centered at the initial value, usually 2 or 3 times the $\Delta x$ step. These two sanity checks are considered when the root occurs in strong nonlinear regions or, more frequently, when the roots are too close to local minima/maxima. In these special situations, the root cannot be computed and a flag index is returned along with the initial value, $x_{rr0}$, as root's best estimation.

\subsection{Example of application}\label{subsec:example1d}

\begin{figure}[!h]
	\centering
	\includegraphics[width=0.90\textwidth]{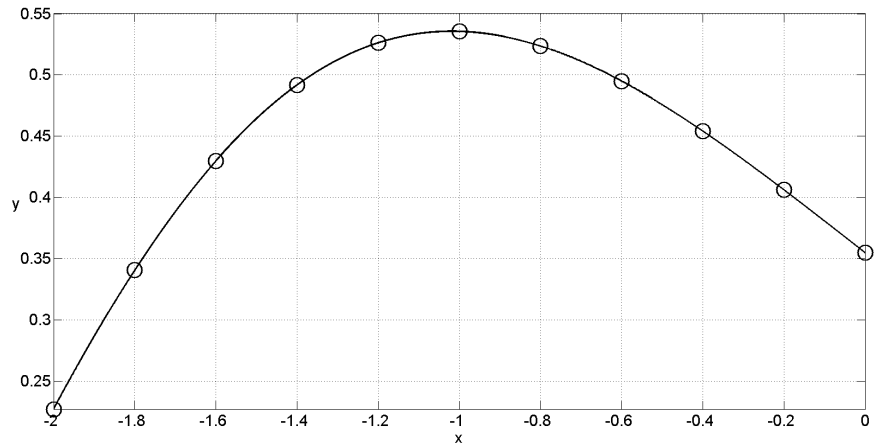}
	\caption{Airy function of the first kind.}
	\label{fig:airy}
\end{figure}

In order to make the procedure clear, consider the Airy function of the first kind,
\begin{equation}
    y = \dfrac{1}{\pi} \int_0^{\infty} \cos\left(\dfrac{t^3}{3} + xt\right) \, \text{d} t,
\end{equation}
with a domain of interest in $x \in \left[-2, 0\right]$. This function is shown in Fig.~\ref{fig:airy}. Suppose that only $n = 11$ points are considered\footnote{This is an extreme case since the \kv\ technique is devised for large databases.}.

From these initial conditions, we first generate the $\B{y}$ database using an uniform distribution of points in $x$. Then, $\B{y}$ is sorted, obtaining the sorted database, $\B{s}$, and the sorting index vector, $\B{I}$. Using this data, the values of $\delta = 0.1134$, $m = 0.0308$ and $q = 0.1966$ are obtained to compute the 11 values of $\B{y}_l(i)$. Then, the \kv\ , $\B{k}_v$, is computed. These vectors are provided in Table~\ref{tab:f(x)}.
\begin{table}[!h]
	\caption{Generation of the \kv.}
	\label{tab:f(x)}
	\centering
	{\scriptsize
		\begin{tabular}{|c|c|c|c|c|c|c|c|c|c|c|c|}
			\hline
			$i$ & $1$ & $2$  & $3$ & $4$ & $5$ & $6$ & $7$ & $8$ & $9$ & $10$ & $11$ \\
			\hline
			$\B{x}$ & $-2.0$ & $-1.8$  & $-1.6$ & $-1.4$ & $-1.2$ & $-1.0$ & $-0.8$ & $-0.6$ & $-0.4$ & $-0.2$ & $0.0$ \\
			\hline
			$\B{y}$ & $0.227$ & $0.341$ & $0.430$ & $0.492$ & $0.526$ & $0.536$ & $0.524$ & $0.495$ & $0.454$ & $0.406$ & $0.355$ \\
			\hline
			$\B{s}$ & $0.227$ & $0.341$ & $0.355$ & $0.406$ & $0.430$ & $0.454$ & $0.492$ & $0.495$ & $0.524$ & $0.526$ & $0.536$ \\
			\hline
			$\B{I}$ & $1$ & $2$ & $11$ & $10$ & $3$ & $9$ & $4$ & $8$ & $7$ & $5$ & $6$ \\
			\hline
			$\B{y}_l$ & $0.227$ & $0.258$ & $0.289$ & $0.320$ & $0.351$ & $0.382$ & $0.412$ & $0.443$ & $0.474$ & $0.505$ & $0.536$ \\
			\hline
			$\B{k}_v$ & $0$ & $1$ & $1$ & $1$ & $2$ & $3$ & $4$ & $5$ & $6$ & $8$ & $11$ \\
			\hline
		\end{tabular}
	}
\end{table}

\begin{figure}[!h]
	\centering
	\includegraphics[width=0.90\textwidth]{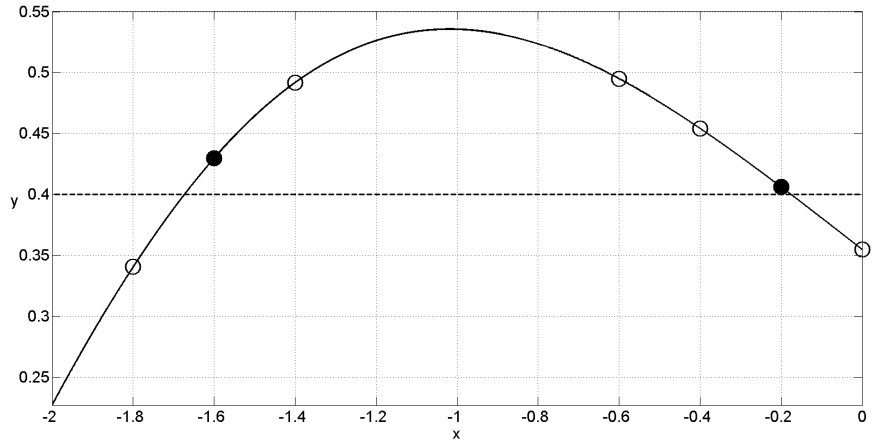}
	\caption{Selected points of the database.}
	\label{fig:airy2}
\end{figure}

At this point the preprocessing is completed and we can select a value of $y_r = f(x)$ and compute the corresponding roots. Let $y_r = 0.4$. Therefore, the searching interval is $[y_a, \: y_b] = [0.2866, \: 0.5134]$\footnote{In this particular example, the interval $[y_a, \: y_b]$ is large due to the lower number of points used, being the searching interval noticeable smaller with larger databases.}. For this range the \kv\ provides the interval indexes $(k_a,k_b)$, which can be applied to the initial set of points in order to find the elements inside the defined interval:
\begin{equation}
    \{\B{x}_s, \: \B{y}_s\} = \{\B{x}(\B{I}_x(\B{I}(k_a:k_b))),\B{y}(\B{I}_x(\B{I}(k_a:k_b)))\},
\end{equation}
where the points $\{\B{x}_s, \: \B{y}_s\}$ are sorted in $x$ in the ascending order, and where $\B{I}_x$ is the sorting vector that relates the sorted and unsorted subset retrieved. These points are shown in Fig.~\ref{fig:airy2} by circle marks (both empty and filled). The numerical results of $\B{k} = \{k_a:k_b\}$, $\B{x}_s$, and $\B{y}_s$, are provided in Table~\ref{tab:interval}.
\begin{table}[!h]
	\caption{Searching interval.}
	\label{tab:interval}
	\centering
	{\scriptsize
		\begin{tabular}{|c|c|c|c|c|c|c|c|}
			\hline
			$i$ & $1$ & $2$  & $3$ & $4$ & $5$ & $6$ & $7$ \\
			\hline
			$\B{k}$ & $2$ & $3$ & $4$ & $5$ & $6$ & $7$ & $8$ \\
			\hline
			$\B{x}_s$ & $-1.8$ & $-1.6$ & $-1.4$ & $-0.6$ & $-0.4$ & $-0.2$ & $0.0$ \\
			\hline
			$\B{y}_s$ & $0.3408$ & $0.4299$ & $0.4917$ & $0.4948$ & $0.4542$ & $0.4063$ & $0.3550$ \\
			\hline
			$\B{I}_x$ & $1$ & $4$ & $6$ & $7$ & $5$ & $3$ & $2$ \\
			\hline
		\end{tabular}
	}
\end{table}

To compute the number of roots, Eqs.~\eqref{nroot1} and~\eqref{nroot2} are used. Then, a search in the subset of points $\{\B{x}_s, \: \B{y}_s\}$ is performed to find the closest point to each root. These are, $x_{r_1}=-1.6$ and $x_{r_2}=-0.2$, providing the values of $\lvert y_r - y_{r_i} \rvert$ of $0.0299$ and $0.0063$, respectively. These points are represented in Fig.~\ref{fig:airy2} as filled circles.

Finally, the Newton-Raphson method is then applied using $x_{r_1}$ and $x_{r_2}$ as starting points, obtaining the roots $x_{r_1} =-1.674$ and $x_{r_2} =-0.17506$, which have an error smaller than $10^{-15}$ after five iterations. As mentioned before, the initial accuracy of the starting points is improved by increasing the number of points in the database. This will reduce the number of iterations as well.

\subsection{Working with tabulated data}\label{sec:clouds}

The \kv\ methodology can be applied even when the function to invert is unknown. In fact, consider a cloud of points that follow a particular unknown function. These points, for example, can represent experimental data, described by an unknown function. In such cases the \kv\ methodology works similarly as presented before. The only difference is that a local interpolation of the points near the roots has to be performed in order to apply the Newton-Raphson method or another root finder.

\section{Big slopes, piecewise defined functions, and singularities}\label{sec:specialcases}

The proposed approach can also be applied to functions with singularities, big slopes, and piecewise defined functions. The following subsections explain how to approach these three cases.

\subsection{Big slopes}\label{subsec:peaks}

For functions characterized by big and small values of their first derivatives, a uniform distribution of $n$ points in $x$ yields to a large value of $\delta$. This implies that the searching interval increases, which makes the methodology running slower. Two possible solutions exist to increase the performance in these cases.
\begin{itemize}
\item If the regions where the function present higher derivatives are well defined, it is possible to separate the $x$ domain in sub-intervals and then, generate different \kv\ for each sub-interval. This way, the roots must be searched in each sub-interval. This procedure allows to adapt the \kv\ to the requirements of each function, increasing the performance of the method.
\item The other methodology consists of generating a non uniform distribution of points in $x$ in such a way that more elements are located where the function presents bigger derivative values. A nonuniform distribution can take advantage of the properties of the particular function and, as it will be seen later, it is also the basic concept behind the ``optimal'' \kv\ methodology.
\end{itemize}

\subsection{Piecewise defined functions}\label{subsec:jumps}

If piecewise defined functions are used, the \kv\ can still be applied, even if the function presents discontinuities. In order to overcome this issue, an initial distribution of points over $x$ is required to be defined in each of the intervals of the function and then, a merging is performed in order to build the final distribution of points. This generates, in general, a non-uniform distribution of points however, this proposed \kv\ approach works with no additional modifications.

\subsection{Singularities}\label{subsec:singularities}

Functions with singularities implies two major issues. First, the regions near the singularities must be avoided to bound the values of $\B{y}_l (i)$. In these singular regions the methodology cannot be applied. Therefore, a range of interest, $[y_{\min}, y_{\max}]$, must be set that bounds the $y_r$ query. This makes the function piecewise (see Section~\ref{subsec:jumps}). Second, due to the high derivative values near the singularity, the searching interval ($\delta$) increases and, consequently, the number of retrieved elements also increases and the process slows down. Again, in order to avoid this, a non uniform distribution of the points in the database is the solution (see Section~\ref{subsec:peaks}).

\subsection{Example of application}\label{subsec:gamma}

\begin{figure}[!h]
	\centering
	{\includegraphics[width=0.90\textwidth]{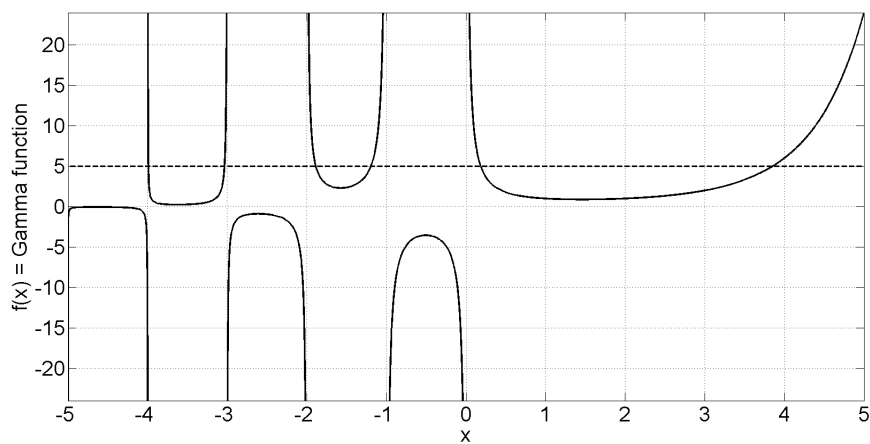}}
	\caption{The $\Gamma(x)$ function}
	\label{fig:gamma}
\end{figure}

To show the procedure in a function with singularities, the $\Gamma(x)$ function has been selected. This function (see Fig.~\ref{fig:gamma}) is defined as
\begin{equation}
    y = \Gamma (x) = \int_0^{\infty}t^{x-1}e^{-t}dt,
\end{equation}
which has singularities in $\mathbb{Z}$ when $\mathbb{Z} \leq 0$.

Without losing generality, we consider the roots in the interval $x \in [-5, \: 5]$ and in the range $y \in [-24, \: 24]$. These two intervals define our region of study. However, we need to define the domain of the region of interest. To do that, we select two values slightly outside the desired range, for example $y_1 =-24.1$ and $y_2 = 24.1$, and we compute the roots for those values. This calculation is done just once and can be performed by any method, for example using the general procedure of the \kv\ methodology.

With the knowledge of these roots, we can define a series of intervals whose union defines the region of interest. In this example, the intervals computed are:
\begin{eqnarray}
x & \in & [-4.99965, \: -4.00172]\cup[-3.99827, \: -3.00686]\cup \nonumber \\
& \cup & [-2.99302, \: -2.02037]\cup[-1.97883, \: -1.04087]\cup \nonumber \\ 
& \cup & [-0.95766, \: -0.04259]\cup[0.04059, \: 5.00276].
\end{eqnarray}
Now, for each of these intervals, a \kv\ can be defined, considering this example as a piecewise function definition. However, the issue with the high values of the derivative still remains and, as explained in Section~\ref{subsec:peaks}, a non-uniform distribution of points in each interval should be introduced.

As it can be seen in Fig.~\ref{fig:gamma}, the function is very abrupt near the extremes of each interval, and thus, more points are required in these regions. In order to take advantage of this fact, we use a distribution of points for each interval following a function that distributes more points near the boundaries of the intervals. An example of this kind of functions is:
\begin{equation} \label{eq:distributiongamma}
    \B{x}(i) = x_{\min} + \dfrac{1}{2} (x_{\max} - x_{\min})\left[1 + \tanh\left(5\pi \, \dfrac{2i - n - 1}{n - 1}\right)\right],
\end{equation}
where $x_{\min}$ and $x_{\max}$ are the extremes of each interval, $n$ is the number of points distributed in the interval and $i\in\{1,\cdots,n\}$ is the distribution variable of these points. Distinct \kv s are built for each one of these intervals; specifically, six \kv s are computed. This is done as part of the preprocessing effort, and therefore, they do not affect the performance of the method while computing the inversions of the function.

Once the \kv s are built, each time that a root is required to be computed, a different search in each of the intervals is performed, obtaining a set of roots for each interval that correspond to the roots of the function. Note that depending on the value of $y_r$ there might be intervals containing no root, and thus, the associated \kv\ does not retrieve any element. As an example of inversion, if $y_r = 5$, the inverse of the function provides six roots,
\begin{equation}
    \mathbf{x}_r = \left\{-3.99156, \: -3.03207, \: -1.88692, \: -1.19389, \: 0.18449, \: 3.85236 \right\},
\end{equation}
with $10^{-14}$ accuracy. This example was performed with $100$ points per interval with five Newton-Raphson iterations. It is important to note that instead of using Eq.~\eqref{eq:distributiongamma}, the optimal \kv\ methodology can be applied obtaining a better performance at a cost of a longer preprocessing.

\section{Optimal \kv\ for function inversion}\label{sec:optimal}

The ``optimal \kv'' for function inversion is an optimization of the former methodology. The improvement is based on the generation of a database distribution, $[\B{x}, \B{y}]$, in such a way that the number of elements retrieved per root is constant for any $y_r$ value in the $[y_{\min}, y_{\max}]$ searching range of interest. In order to obtain this feature, an additional preprocessing to generate the new point distribution is required. However, once the preprocessing is done, the function inversion becomes extremely fast as the number of points retrieved per root is optimal for the root finder selected: one for Newton, two for bisection or regula-falsi, or more, if required.

Building the optimal \kv\ requires three steps:
\begin{enumerate}
\item The \kv\ is generated using an uniform distribution of points. This \kv\ is then used to calculate the roots by following the methodology previously described.
\item The new distribution of points is computed. Let $n_d$ be the number of levels defined in the range of the function. This value affects the size of the optimal \kv\ that we want to define (number of elements). The parameter $n_d$ is free to choose. Larger values of $n_d$ increases the memory required but also makes the root solver to run faster since the points retrieved will be closer to the roots (less iterations). This means that \emph{this methodology can be customized for performance and memory requirements}. Let $\B{y}_d$ be a linear distribution over the range of the function, that is,
    \begin{equation}\label{y_dis}
        \B{y}_d (i) = y_{\min} + \dfrac{y_{\max} - y_{\min}}{n_d-1}(i-1),
    \end{equation}
    where $i = {1, 2,\cdots, n_d}$. Let $\{\B{x}_d(i)\}$ be the set of roots for each value of $\B{y}_d(i)$ computed using the \kv\ methodology for function inversion. Then, the new distribution of points is given by the set consisting on all the roots computed, that is:
    \begin{equation}
        \B{x} = \left\{\bigcup_{i=1}^{n_d}\{\B{x}_d(i)\}\right\},
    \end{equation}
    where the distribution $\B{x}$, along with its images $\B{y} = f(\B{x})$, substitutes the initial database stored in memory.
\item The \kv\ is built for the new database, $\B{x}$. However, this new \kv\ has an important feature: for an assigned searching range
    \begin{equation}
        [y_a, \; y_b] = \left[y_r - n_e\dfrac{\delta}{2}, \; y_r + n_e \dfrac{\delta}{2}\right] \quad \text{where} \quad \delta = \dfrac{y_{\max} - y_{\min}}{n_d - 1} + 4 \varepsilon,
    \end{equation}
    \emph{the number of elements retrieved is always the same}. $y_r$ indicates the value of the function to invert and $n_e$ the number of elements associated with each root. That way, if $n_e = 1$, only one element per root is obtained, which corresponds to the closest point of the database to the root. If $n_e = 2$, then two elements are always retrieved per root, which are the two closest elements of the database to the roots and, in addition, these \emph{two elements bracket the root}. This has important implications: 1) no additional search is required in order to find the closest point(s) from the database to the roots, 2) the points retrieved (and the closest in the database) constitute a sanity check bound for Newton-Raphson root solver iterations.
\end{enumerate}

\subsection{Example of application}

\begin{figure}[!h]
	\centering\includegraphics[width=0.90\textwidth]{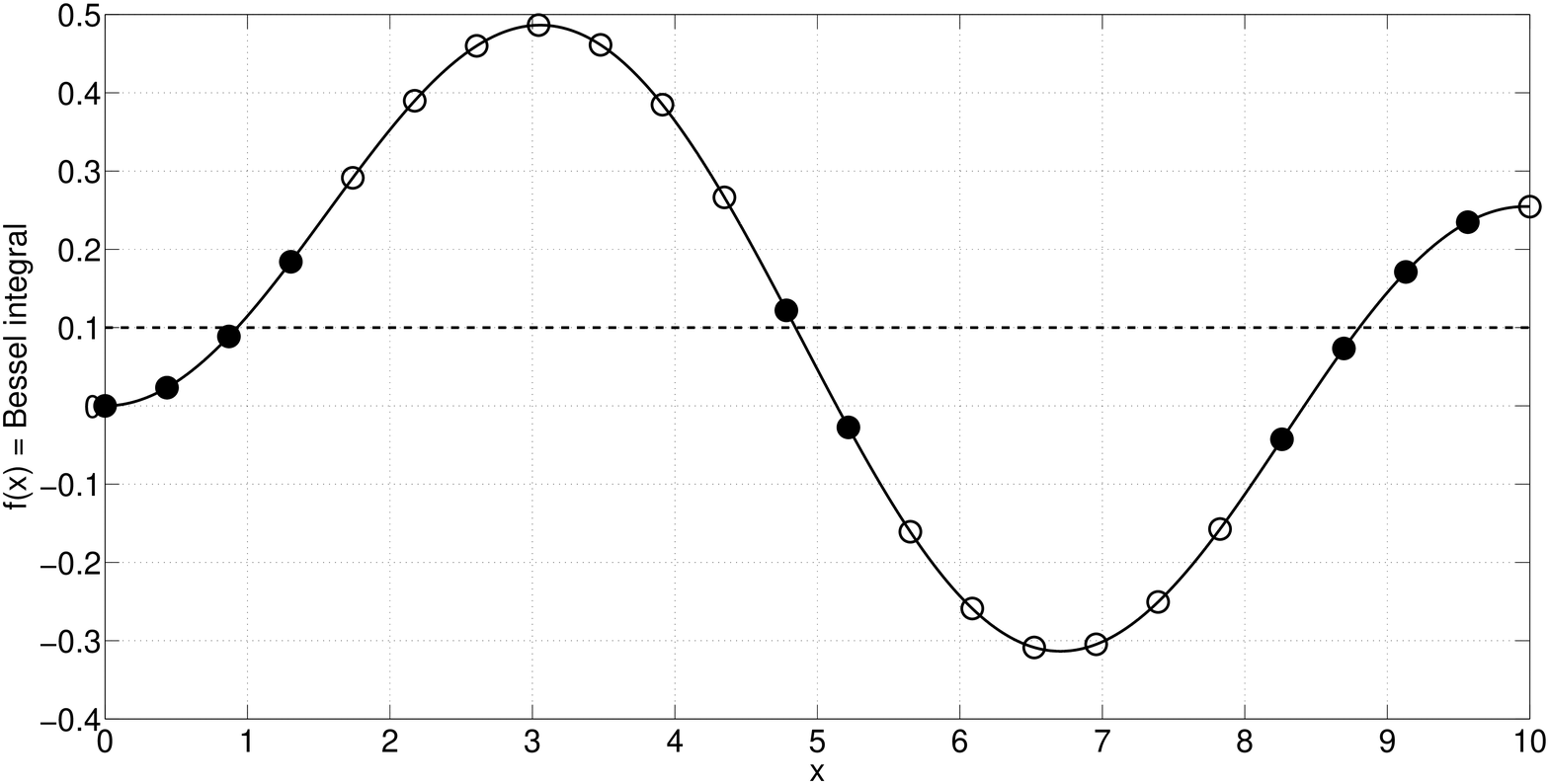}
	\caption{General procedure for Bessel's integral.}
	\label{fig:bessel2}
\end{figure}

As an example of application, let $y = f(x)$ be the Bessel integral defined by:
\begin{equation}
    y = \ds\dfrac{1}{\pi}\int_{0}^{\pi} \cos(2t - x\sin t) \, dt,
\end{equation}
and suppose the roots contained in $x\in [0,10]$ are the ones of interest. To show a clear example, a small database of 24 points is selected. Figure~\ref{fig:bessel2} shows the Bessel integral and the 24 points used to generate the initial \kv\ (represented by empty and filled circles).

\begin{figure}[!h]
	\centering\includegraphics[width=0.90\textwidth]{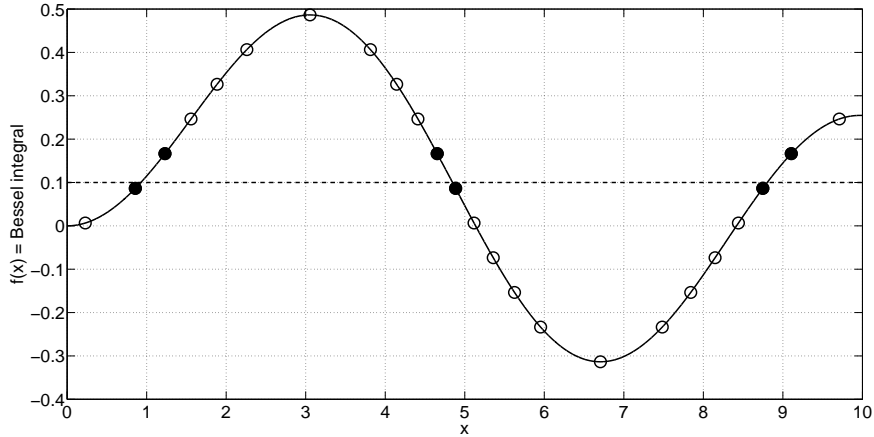}
	\caption{Improved search for Bessel's integral.}
	\label{fig:bessel1}
\end{figure}

Now, we define the levels of the function with Eq.~\eqref{y_dis} and compute the roots related to them using the \kv\ for function inversion. The roots obtained are now used as the new distribution of points in order to build the optimal \kv. Table~\ref{tab:gauss} shows the results for this example when $n_d = 11$ levels are defined. This distribution can also be seen in Fig.~\ref{fig:bessel1}, where the circle marks (filled and emptied) represent the points of the distribution used in the optimal \kv.

\begin{table}[!h]
	\caption{New distribution of points.}
	\label{tab:gauss}
	\centering
	{\scriptsize
		\begin{tabular}{|c|c|c|c|c|c|c|}
			\hline
			$\B{y}$ & $0.0065$ & $0.0865$ & $0.1665$ & $0.2465$ & $0.3265$ & $0.4065$ \\
			\hline
			$\B{x}$ & $0.2282$ & $0.8579$ & $1.2305$ & $1.5579$ & $1.8854$ & $2.2596$ \\
			\hline
			\hline
			$\B{y}$ & $0.4865$ & $0.4065$ & $0.3265$ & $0.2465$ & $0.1665$ & $0.0865$ \\
			\hline
			$\B{x}$ & $3.0543$ & $3.8137$ & $4.1433$ & $4.4113$ & $4.6535$ & $4.8851$ \\
			\hline
			\hline
			$\B{y}$ & $0.0065$ & $-0.0735$ & $-0.1535$ & $-0.2335$ & $-0.3135$ & $-0.2335$ \\
			\hline
			$\B{x}$ & $5.1166$ & $5.3583$ & $5.6256$ & $5.9536$ & $6.7061$ & $7.4833$ \\
			\hline
			\hline
			$\B{y}$ & $-0.1535$ & $-0.0735$ & $0.0065$ & $0.0865$ & $0.1665$ & $0.2465$ \\
			\hline
			$\B{x}$ & $7.8417$ & $8.1476$ & $8.4412$ & $8.7478$ & $9.1060$ & $9.7100$ \\
			\hline
		\end{tabular}
	}
\end{table}

Once the preprocessing is done, the calculation of the roots can be made using the general procedure. As an example, we compute the roots for $y_r = 0.1$. The proposed methodology provides the roots with machine error precision, $x_r = \{0.9274, 4.8462, 8.8031\}$, using the elements retrieved by the optimal \kv\ and just five Newton-Raphson iterations.

Figure~\ref{fig:bessel1} also shows that the number of points per root retrieved is $n_e = 2$, with all roots bracketed by the elements retrieved. We can compare this result with the general case shown in Fig.~\ref{fig:bessel2}, which retrieved $10$ points. This effect is more evident if only $n_e = 1$ point per root is selected as, for instance, when using a Newton-Rahpson root finder.

\section{Optimal \kv\ with no function evaluation}

The optimal \kv\ methodology can also be used as an approximation to function inversion without requiring to evaluate the function during the inversion process. This is specially useful when very fast inversions of the same function are required and more memory is available. One of the applications of this methodology is to generate a function inversion toolbox or a random number generator following a particular distribution function~\cite{random}.

In Section~\ref{sec:optimal}, the optimal \kv\ was introduced, providing a general methodology to always retrieve the same number of elements from the database. This technique removed the requirement of performing searches inside the elements retrieved, and was also able to bracket all the roots of the function. The objective now is to improve the speed performance of the methodology by substituting the final root finder iterations with an approximating expression using a larger database containing information about the derivatives of the function in points of interest. In that respect, two different methodologies are introduced using a linear approximation, and Householder's methods (in particular, Newton-Raphson for a first order method, and Halley for a second order approximation). These methodologies do not require to perform any searches (as a characteristic of the optimal \kv) nor to evaluate the function (since all the information is provided by the database) during the inverse process. This allows to invert functions in a very fast way, using the stored points and derivatives of the function, which were created in a single more intensive preprocessing.

\subsection{Optimal \kv\ with linear approximation}

Let $y_r$ be the value of the function whose roots are required to be computed. Then, by setting $n_e = 2$, the optimal \kv\ retrieves two elements per root for each value of $y_r$. Let $\B{x}_s$ be the elements retrieved from the database sorted in the ascending mode in $x$, and let $\B{I}_x$ be the sorting index vector. That way:
\begin{equation}
    \{\B{x}_s, \: \B{y}_s\} = \{\B{x}(\B{I}_x(\B{I}(k_a:k_b))),\B{y}(\B{I}_x(\B{I}(k_a:k_b)))\}.
\end{equation}
Moreover, let $n_r$ be the number of roots, then, the size of $\B{x}_s$ and $\B{y}_s$ is $2n_r$. This means that each consecutive two points correspond to a root, and thus, an approximation of the root can be performed by a linear interpolation between the two points, that is:
\begin{equation}
    \B{x}_r(i) = \B{x}_s(2i-1) + \ds\dfrac{y_r - \B{y}_s(2i-1) }{\B{y}_s(2i) - \B{y}_s(2i-1) }(\B{x}_s(2i) - \B{x}_s(2i-1) );
\end{equation}
where $i \in \{1,\cdots,n_r\}$ names each root of the function and $\B{x}_r(i)$ are the approximations to the roots of the function in $y_r$.

\subsection{Optimal \kv\ using a first order approximation}

This technique is based on the idea of obtaining the approximation to the root by performing a first iteration of the Newton-Raphson method without requiring to evaluate the function nor its derivative during the inversion. This is done by adding the information of the derivative in the database.

Suppose that a \kv\ has been computed alongside with its database $\B{x}$. As part of the preprocessing, the function $\B{y}$ and its first derivative $\dot{\B{y}}$ in the points $\B{x}$ is computed and stored in memory. Then, by setting $n_e = 1$ as parameter for the searching range, just one point per root is retrieved for each value of $y_r$. In particular, the points $\{\B{x}(\B{I}_x(\B{I}(\B{k})))\}$ retrieved correspond to the closest points of the database to the roots.

Let $n_r$ be the number of roots of the function. Then, the approximation of the root can be computed as:
\begin{equation}
    \B{x}_r(i) = \B{x}(t) - \dfrac{\B{y}(t) - y_r}{\dot{\B{y}}(t)}, \quad \text{where} \quad t = \B{I}_x(\B{I}(\B{k}(i))),
\end{equation}
and $i \in \{1,\cdots,n_r\}$ names each root. Note that the values of $\B{x}(t)$, $\B{y}(t)$ and $\dot{\B{y}}(t)$ used are elements of the database and are not required to be computed.

However, it is important to note that, as the Newton-Rahpson method is based in a linearization of the function, the error provided by this methodology is similar to the linear interpolation for large databases. Therefore, it is preferable in general to use the linear interpolation as it does not require to increase the database size.

\subsection{Optimal \kv\ using a second order approximation}

The accuracy of the first order approximation can be improved using the Halley's method~\cite{halley}. This methodology requires that, in addition to the preprocessing performed in the first order approximation, the second derivative in the points of the distribution has to be computed and stored in memory ($\ddot{\B{y}}$). The process is similar to the first order approximation. First, we retrieve the elements $\B{x}(\B{I}_x(\B{I}(\B{k}(i))))$. Then, the second order approximation is performed:
\begin{equation}
    \B{x}_r(i) = \B{x}(t) - \ds\dfrac{2(\B{y}(t) - y_r)\dot{\B{y}}(t)}{2\dot{\B{y}}(t)^2 - (\B{y}(t) - y_r)\ddot{\B{y}}(t)}, \quad \text{where} \quad t = \B{I}_x(\B{I}(\B{k}(i))),
\end{equation}
and $i \in \{1,\cdots,n_r\}$ names each root, and the values of $\B{x}(t)$, $\B{y}(t)$, $\dot{\B{y}}(t)$ and $\ddot{\B{y}}(t)$ used are elements of the database. This methodology provides more accurate values of the roots with respect to the first order approximation at the cost of a larger database.

It is important to note that it is possible to increase the order of the methodology by storing higher order derivatives of the function and performing a higher order interpolation by Taylor series and using Householder's methods~\cite{Householder}. This increases the memory requirements and the time to compute the inverse, nevertheless, the accuracy of the methodology is improved. Another possibility to improve the accuracy without increasing the run time is to increase the size of the database for the \kv. That way, the initial approximation retrieved from the database is closer to the real roots of the function.

\subsection{Example of application}

As an example of application of the optimal \kv\ with no function evaluation, Kepler's equation is studied, which has the following form:
\begin{equation}
    y = x - e\sin x;
\end{equation}
where $e$ is the parameter of eccentricity of a conic. Without losing generality $e = 0.5$ is selected as parameter from the family of functions given by the Kepler's equation. A large value of $e$ is selected due to the increase in the nonlinearity of the function with this parameter. Then, an optimal \kv\ based on $n = 65,535$ elements is generated with its respective databases: $\B{x}$, $\B{y}$, $\dot{\B{y}}$, and $\ddot{\B{y}}$.

First, an approximation error analysis is performed by finding the maximum error that can be produced by the approximation. In that respect, we consider the middle points:
\begin{equation}
    \B{x}_r(i) = \ds\dfrac{\B{x}(i) + \B{x}(i+1)}{2}, \quad \text{with} \quad i = \{1,\cdots,n-1\},
\end{equation}
for function evaluation, that is:
\begin{equation}
    \B{y}_r(i) = \B{x}_r(i) - e\sin \B{x}_r(i), \quad \text{with} \quad i = \{1,\cdots,n-1\},
\end{equation}
where $\B{y}_r(i)$ are the values of the function where to compute the inverse.

\begin{figure}[!h]
	\centering
	{\includegraphics[width=0.90\textwidth]{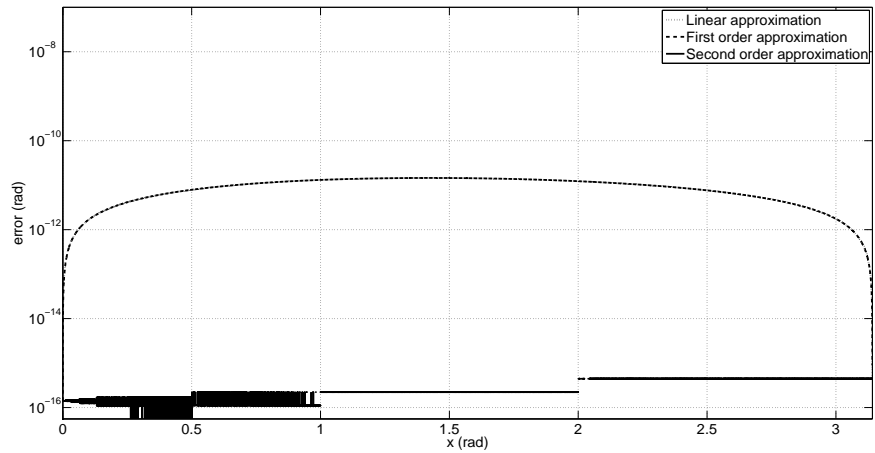}}
	\caption{Error obtained in Kepler's equation for $e=0.5$.}
	\label{fig:kepler05}
\end{figure}

Figure~\ref{fig:kepler05} shows the error of the three methodologies, using linear, first order, and second order approximations. It is interesting to note that the first order approach and the linear approximation provide a very similar accuracy due to the small distance between points (in fact, in the figure cannot be distinguished). This means that in general, it is more convenient to use the linear interpolation, as it does not require to increase the size of the database while maintaining the accuracy. On the other hand, the second order approximation provides the best accuracy as expected. As it can be seen, with a second order approximation we can obtain near machine error precision in all the range. Regarding the linear interpolation, it presents a maximum error of $4 \cdot 10^{-10}$ radians, 5 orders of magnitude better than the original points retrieved from the database. Remember that these results were obtained without performing an iteration nor a function evaluation.

\begin{figure}[!h]
	\centering
	{\includegraphics[width=0.90\textwidth]{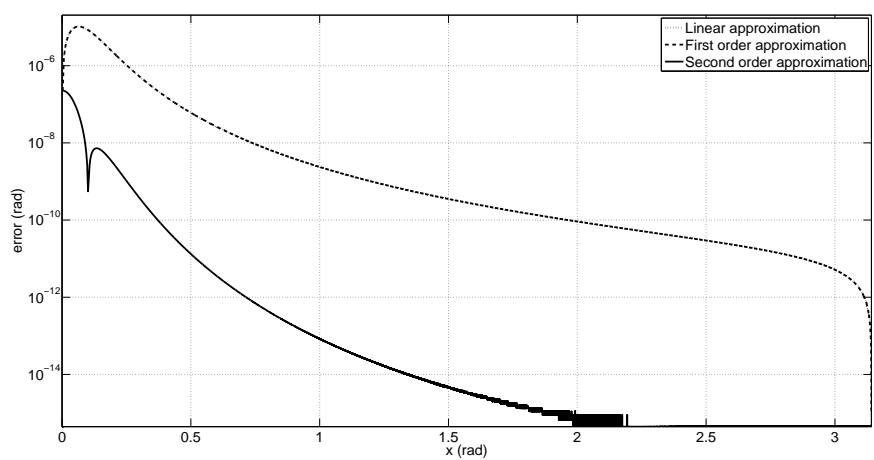}}
	\caption{Error obtained in Kepler's equation for $e=0.99$.}
	\label{fig:kepler99}
\end{figure}

On the other hand, another example with Kepler's equation is provided in order to show the behavior of this methodology in the singular corner of the equation. The singular corner of Kepler's equation is the region where the eccentricity is near one and $x$ presents values near zero. In that region, all numerical methods require more iterations to converge due to the high nonlinearity of the equation in that region~\cite{Conway,Elipe,Fukushima,Mikkola}. Thus, in order to test the algorithm in this region, a value of $e=0.99$ has been selected. The errors of the different methodologies can be seen in Fig.~\ref{fig:kepler99}. As it can be observed, the error has increased with respect to the other example. Nevertheless, it is possible to obtain a maximum error of $10^{-7}$ rad for all the range using a second order approximation and without evaluating the function. As mentioned earlier, this accuracy can be improved if required by increasing the order of the method or by continuing the iteration with a root finder.

\section{Performance of the methodology}

In this section we deal with the performance of the \kv\ methodology in terms of speed. In order to do that, several numerical test are performed in a single threaded
script that was coded in Matlab on an Intel i7-3770 processor running at 3.4 GHz, with a total memory of 8 GB of RAM (which means that only 1 GB is on actual use) and Windows 7. In addition, the Gaussian integral (which is related to the error function) is selected as the objective function. This integral is defined by:
\begin{equation}
y = \int_{-\infty}^{x}\frac{1}{\sigma\sqrt{2\pi}}e^{-\frac{\left(t-\mu\right)^2}{2\sigma^2}} \, \: \text{d} t,
\end{equation}
where $\sigma = 0.2$ and $\mu = 0$ are chosen as parameters of the function. Moreover, and without losing generality, we assume that only the roots between $x\in [-1,1]$ are considered.

Under these conditions, we compare the performance of the \kv\ with a database of $1,000$ points, the optimal \kv\ without function evaluation and the Matlab \verb"fzero" function~\cite{Brent, Forsythe} which is based on a Newton-Raphson method. In the case of the optimal \kv, a database composed by $1,000$ points and their derivatives up to 4th order is stored in memory, since a fourth order approximation will be used. In order to perform these speed tests, $10$ different runs for each algorithm considered are performed, each one consisting in $100,000$ function inversions of a random value in the range $y\in[0,1]$ with a prescribed accuracy of $10^{-15}$.

\begin{table}[!h]
	\caption{Speed performance of the algorithms for $100,000$ function inversions.}
	\label{tab:performance}
	\centering
	{\scriptsize
		\begin{tabular}{|c|c|c|c|}
			\hline
			Algorithm & \verb"fzero" & \kv & optimal \kv \\
			\hline
			Time spent (s) & $72.3$ & $27.2$ & $1.8$ \\
			\hline
			Mean number of iterations & $9.75$ & $1.62$ & $0$ \\
			\hline
		\end{tabular}
	}
\end{table}

Table~\ref{tab:performance} shows the mean time and the mean number of iterations that each algorithm required to perform the inversions. As it can be seen, the \kv\ methodology is able to perform the inversion faster and with a lower number of iterations than a general Newton-Raphson method, since it provides a closer starting point for the initialization of the iteration. On the other hand, the optimal \kv\ does not require any iteration, providing the root with just the original database. This allows to reduce the time required to perform the inversion in more than one order of magnitude with respect to the other algorithms. However, the methodology requires a larger database (4 derivatives have been stored) and, although we obtain machine error precision in nearly all the range, there is a very small area near the extremes of the domain (that is, close to $x=-1$ and $x=1$) where the precision is reduced up to $10^{-8}$. This situation is provoked by the relative lack of points distributed in those areas (compared to the rest of the function) and the increase of the nonlinearity. This issue can be easily solved by generation an additional \kv\ for those areas. That way, machine error precision is obtain in all the domain.

\section{Conclusions}\label{sec:conclusions}

Finding roots of a function is one of the most common tasks in science. When the solution presents an analytical expression, it is straightforward to evaluate the roots and the behavior of the function in study. However, there are cases in which there is no known analytical solution and thus, numerical methods are required in order to solve the function. In this work we introduce the \kv\ as a general methodology to invert any kind of function.

The \kv\ is a range searching technique originally developed for the problem of star identification in star trackers. The methodology works with static databases and requires an initial set up. However, once this preprocessing is performed, the \kv\ is a nearly search-less algorithm, which makes it specially interesting for large databases. This idea is applied in this work to function inversion, by generating a database with the values of the function, and then, performing fast searches in the database created in order to find root approximations of the function. Finally, a root finder is applied to obtain a prescribed accuracy.

The methodology has proved to be extremely fast for recursive function inversion and it has been applied to a wide variety of functions with good results, being able to obtain all the roots of the function at once. Examples of that are Airy functions, Dawson's integrals, Bessel's integrals, Kepler's equation, elliptic integrals, Fresnel's integrals or polygamma functions. Moreover, the performance of the methodology can be improved even further by the use of the optimal \kv\, which makes the process of bracketing the roots completely search-less. The optimal \kv\ consist of finding the distribution of points such that each time that the \kv\ is used, the same number of elements per root is always retrieved. In particular, these elements correspond to the closest points to the roots. This modification of the \kv\ technique requires an additional preprocessing, but once this process is done, the robustness of the methodology is enhanced considerably.

In addition, the optimal \kv\ technique can also be used in combination with a linear or a Householder approximation. This allows to obtain a root approximation without the requirement of evaluating the function in the process (just in the preprocessing). That way, the inversion process becomes even faster, as the methodology does not require to spend resources in the computation of the function nor its derivatives. However, this procedure requires to store a larger database and its accuracy depends on the size of the database, the order of the approximation and the function studied.

On the other hand, the \kv\ technique (and its optimal version) can adapt its performance to the memory available, and the speed and accuracy required. That way, if the memory available is low, a lower dimension \kv\ can be built. On the other hand, if more speed and accuracy are required, larger databases can be used in order to speed the process. This provides tools for adapting the methodology to the function to solve and the resources available.

Finally, the complexity of the algorithm is not too high, making the \kv\ an easy to implement methodology with a wide range of applications. Moreover, the initial distribution of the database can be adapted to the function in study, providing the possibility to optimize the method for particular functions in high demanding applications.

\section{Acknowledgments}
The work of D. Arnas was supported by the Spanish Ministry of Economy and Competitiveness (Project no. ESP2013\textendash44217\textendash R) and the Research Group E48: GME.


\clearpage
\appendix{APPENDIX A: \kv\ preprocessing}
\label{appendix1}

Algorithm to build the \kv\ from a sorted database:

\begin{algorithm}[H]
	$\mathbf{function} \: [m, q, kv] = \mathbf{BuildKV}(S)$ \\
	$n = length(S)$; \\
	$d = (S(n) - S(1))eps$; \\
	$m = (S(n) - S(1) + 2d)/(n - 1)$; \\
	$q = S(1) - d - m$; \\
	$Yl = (1:n)m + q$; \\
	$i = 1$; \\
	\For {$j$ from $2$ to $(n-1)$} {
		\While {$S(i) < Yl(j)$} {
			$i = i + 1$; \\
		}
		$kv(j) = i - 1$; \\
	}
	$kv(n) = n$; \\
\end{algorithm}

\vspace{1.0cm}

\appendix{APPENDIX B: \kv\ use}
\label{appendix2}

To use the \kv\ the following algorithm can be used:

\begin{algorithm}[H]
	$\mathbf{function} \: k = \mathbf{UseKV}(S, Smin, Smax, m, q, kv)$ \\
	$n = length(S)$; \\
	$i1 = floor((Smin - q)/m)$; \\
	\lIf{$i1<1$}{$i1=1$}
	\lIf{$i1>n$}{$i1=n$}
	$i2 = ceil((Smax - q)/m)$; \\
	\lIf{$i2<1$}{$i2=1$}
	\lIf{$i2>n$}{$i2=n$}	
	$k1f = kv(i1) + 1$; \\
	$k2f = kv(i2)$; \\
	\eIf{$k2f < k1f$}{$k = [\:]$}{
		\For{$j$ from $k1f$ to $k2f$}{
			\If{$S(j) >= Smin$}{break}}
		$k = k1:k$;}
\end{algorithm}

\end{document}